\documentclass[10pt,twocolumn]{icfd2023paper}
\usepackage[dvipdfmx]{graphicx}
\usepackage{ascmac}
\usepackage{amsmath,amssymb}
\usepackage{indentfirst}

\usepackage{ulem}
\usepackage{xcolor,soul}
\usepackage{mathptmx}
\usepackage{color}%
\usepackage{ccaption}
\usepackage{caption}
\usepackage{cite}

\title{Finsler Geometry Modeling and Monte Carlo Study on Geometrically Confined skyrmions in Nanodots}
\author{Gildas Diguet${}^1$,
Benjamin Ducharne${}^{2,3}$, Sahbi El Hog${}^{4}$, Fumitake Kato${}^5$, Hiroshi Koibuchi${}^5$, \\Tetsuya Uchimoto${}^{3,6}$, Hung The Diep${}^7$   
\footnote{\normalsize Corresponding author: Hiroshi Koibuchi}
\footnote{\normalsize {\textit{E-mail address}}:  koi-hiro@sendai-nct.ac.jp}
\\
${}^1$ Micro System Integration Center, Tohoku University, Sendai, Japan \\
${}^2$ INSA Lyon, Universite de Lyon, Villeurbanne Cedex, France \\
${}^3$ ELyTMaX, CNRS-Universite de Lyon-Tohoku University, Sendai, Japan \\
${}^4$ Universit$\acute{\rm e}$ de Monastir (LMCN), Monastir, Tunisie \\
${}^5$ National Institute of Technology (KOSEN), Ibaraki College, Hitachinaka, Japan \\
${}^6$ Institute of Fluid Science, Tohoku University, Sendai, Japan \\
${}^7$ CY Cergy Paris University, Cergy-Pontoise, France\\
}
\abstract{
Using the Finsler geometry modeling (FG) technique without spontaneous magnetic anisotropy, we numerically study the stability and morphology of geometrically confined skyrmions  experimentally observed in nanodots. We find a confinement effect that stabilizes skyrmions for a low external magnetic field without mechanical stresses by decreasing the diameter of the cylindrical lattice and strain effects that cause the sky and vortex to emerge under the zero magnetic field.  Moreover, the obtained MC data on the morphological changes are also consistent with the reported experimental data.
}
\begin{document}
\captionsetup{labelsep=space}

\maketitle
\thispagestyle{empty}

\section{Introduction}
The stability of skyrmion (sky) configurations in chiral magnets and materials hosting skys play key roles in future skyrmion control technology \cite{Fert-etal-NatReview2017,Zhang-etal-JPhys2020,Gobel-etal-PhysRep2021}. The well-known conditions for sky stabilization are external magnetic field, magnetic anisotropy, and magnetoelastic coupling \cite{Bogdanov-PRL2001,Butenko-etal-PRB2010,Seki-etal-PRB2017}. Geometric confinement (GC) was also proposed as a stabilization technique \cite{Rohart-Thiaville-PRB2013}, and a remarkable GC effect in combination with a strain effect was demonstrated in a recent experiment on nanodot skys \cite{Matsumoto-etal-NanoLett2018,YWang-etal-NatCom2020}. In Ref. \cite{Diguet-etal-JMMM2023}, we proposed a model for GC, in which zero Dzyaloshinskii-Moriya interaction (DMI) is assumed on the surfaces parallel to the magnetic field. Strain effects were also implemented in the model via lattice deformations, causing static strains. However, the induced strains are static and have no positional dependence \cite{Diguet-etal-JMMM2023}.

To explain the experimental results presented in Refs. \cite{YWang-etal-NatCom2020}, we introduce a directional degree of freedom $\vec{\tau}(\in\! S^2/2)$ of in-homogeneous strain at each lattice vertex and assume that the interaction length dynamically depends on the direction of  $\vec{\tau}$ in the framework of Finsler geometry (FG), in sharp contrast to the ordinarily assumed constant Euclidean length. The interactions modified by the FG modeling prescription are ferromagnetic interaction (FMI), DMI, and a second-order ferromagnetic interaction with a magneto-elastic coupling. No explicit magnetic anisotropy is assumed; therefore, the material we study is slightly different from that in \cite{YWang-etal-NatCom2020}.  Remarkably, the assumed interactions in our study are dynamically modified to become anisotropic when $\vec{\tau}$ is aligned in a specific direction. The $\vec{\tau}$ direction is controllable by external stress $\vec{f}$, and the controlled and direction-dependent $\vec{\tau}$ modifies the anisotropic interactions that play a role in direction-dependent magnetoelastic coupling and magnetic anisotropy. Moreover, the GC effect is naturally implemented in the FG modeling technique as a small DMI on the surface compared with the bulk DMI.

\section{Method}
\subsection{3D Cylindrical Lattices and Radial Stresses\label{lattices}}
\begin{figure}[ht]
	\centering{}\includegraphics[width=8.0cm,keepaspectratio,clip]{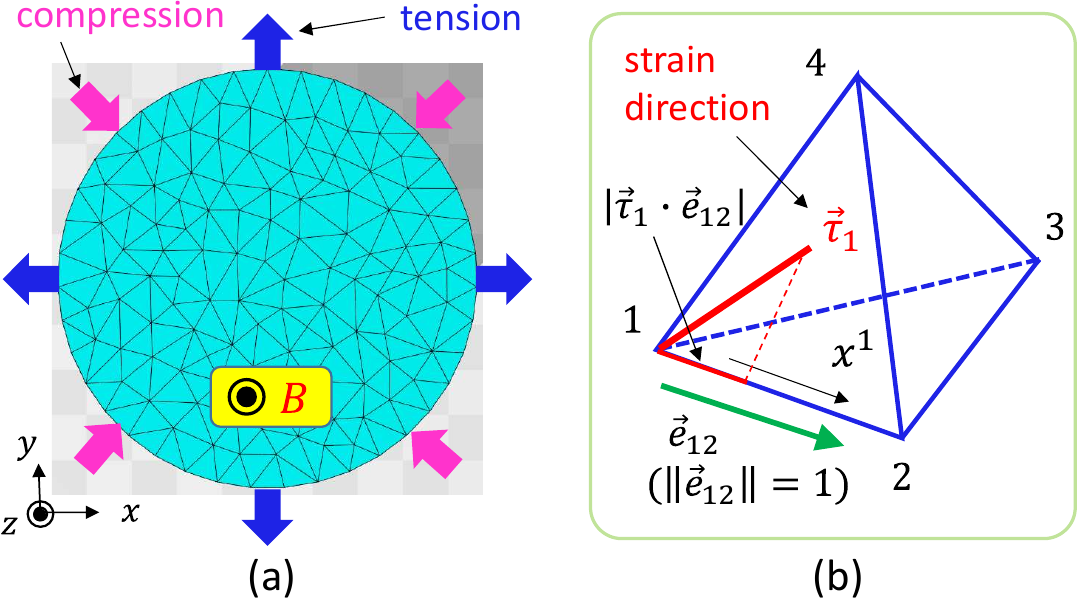}
	\caption{
		(a) Illustrations of tensile and compressive stresses radially applied to a cylindrical lattice composed of tetrahedrons, (b) strain direction $\vec{\tau}_1$ at vertex 1  and its component $|\vec{\tau}_1\cdot\vec{e}_{ij}|$ along a local coordinate axis $x_1$ of a tetrahedron with vertices 1, 2, 3 and 4.   		\label{fig-1} }
\end{figure}
Cylindrical geometry lattices discretized by tetrahedrons are used for the simulations (see Appendix \ref{App-A} for further details of the lattice structure). Mechanical stresses are applied along the radial direction (Fig. \ref{fig-1}(a)), and the magnetic field applied along the $z$ direction. We use three different lattices of size $N\!=\!2083, 5430, 11962$, the total number of vertices, for the ratios $R(=\!D/h)\!=1.2, 2, 3$ of the diameter $D$ with fixed height $h\!=\!12$ (Appendix \ref{App-A}). 
\subsection{Hamiltonian and Monte Carlo\label{Hamiltonian}}
The discrete Hamiltonian is given by 
\begin{eqnarray}
	\label{total-Hamiltonian}
	H(\vec{\sigma},\vec{\tau})=\lambda H_{\rm FM}+DH_{\rm DM}+H_B+H_f+\alpha H_{\rm ME},
\end{eqnarray}
where $\vec{\sigma} (\in S^2)$ and $\vec{\tau} (\in S^2/2)$ denote the spin and strain variables, respectively, defined at each lattice vertex. A free boundary condition is assumed for the variables on the surface. The symbols $\lambda, D$ and $\alpha$ on the right-hand side are the interaction coefficients, and the terms are given as follows:

\begin{eqnarray}
	\label{Hamiltonians}
	\begin{split}
		& H_{\rm FM}=\sum_{\Delta} \sum_{ij(\Delta)}\Gamma_{ij}(\vec{\tau})\left(1-\vec{\sigma}_i\cdot\vec{\sigma}_j\right),\\
		& H_{\rm DM}=\sum_{\Delta}\sum_{ij(\Delta)}\Gamma_{ij}(\vec{\tau})\vec{e}_{ij}\cdot \vec{\sigma}_i\times\vec{\sigma}_j,\\
		& H_B=-\sum_i \vec{\sigma}_i\cdot\vec{B},\quad \vec{B}=(0,0,B), \\  
		& H_f=-{\rm sgn}(f)\sum_{i} \left(\vec{\tau}_i\cdot \vec{f}\right)^2,\\ &\vec{f}=f\frac{\vec{r}}{\|\vec{r}\|},  \quad {\rm sgn}(f)=\left\{ \begin{array}{@{\,}ll}
			1\quad  (f:{\rm tension}) \\
			-1\quad  (f:{\rm compression})
		\end{array} 
		\right., \\
		& H_{\rm ME}={\rm sgn}(f)f\sum_{\Delta} \sum_{ij(\Delta)}\Omega_{ij}(\vec{\tau})\left(\vec{\sigma}_i\cdot\vec{\sigma}_j\right)^2.
	\end{split}
\end{eqnarray}
The first term $H_{\rm FM}$ describes the ferromagnetic interaction (FMI), which is deformed to have a dynamical interaction coefficient $\Gamma_{ij}(\vec{\tau})$ between the nearest neighbors $\sigma_i$ and $\sigma_j$ (see Appendix \ref{App-B} for FG modeling details). Note that $\Gamma_{ij}(\vec{\tau})$ and $\Omega_{ij}(\vec{\tau})$ are normalized, such that $\Gamma_{ij}(\vec{\tau})\!\to\! 1$, $\Omega_{ij}(\vec{\tau})\!\to\! 1$ for isotropic $\vec{\tau}$, which corresponds to the zero-stress configuration (Appendix \ref{App-B}). The second term describes a deformed DMI with the same $\Gamma_{ij}(\vec{\tau})$. The third term is the Zeeman energy and the fourth term is the response energy of strain $\vec{\tau}$ to external stress $\vec{f}$ along the radial direction. Because $\vec{\tau}$ is nonpolar, the factor ${\rm sgn}(f)$ is introduced to distinguish between tensile and compressive stresses.  The final term is the energy for the magnetostriction quadratic with respect to $\vec{\sigma}$ including the coefficient $\Omega_{ij}(\vec{\tau})$, which differs slightly from  $\Gamma_{ij}(\vec{\tau})$ for $H_{\rm FM}$ and $H_{\rm DM}$ (Appendix \ref{App-B}). Note that $H_f\!=\!H_{\rm ME}\!=\!0$ for $f\!=\!0$. To update $\vec{\sigma}$ and $\vec{\tau}$, we use the Metropolis Monte Carlo technique with random initial configurations.

\section {Results and Discussion}
\subsubsection{Confinement Effect}
\begin{figure}[h]
	\centering{}\includegraphics[width=8.0cm,keepaspectratio,clip]{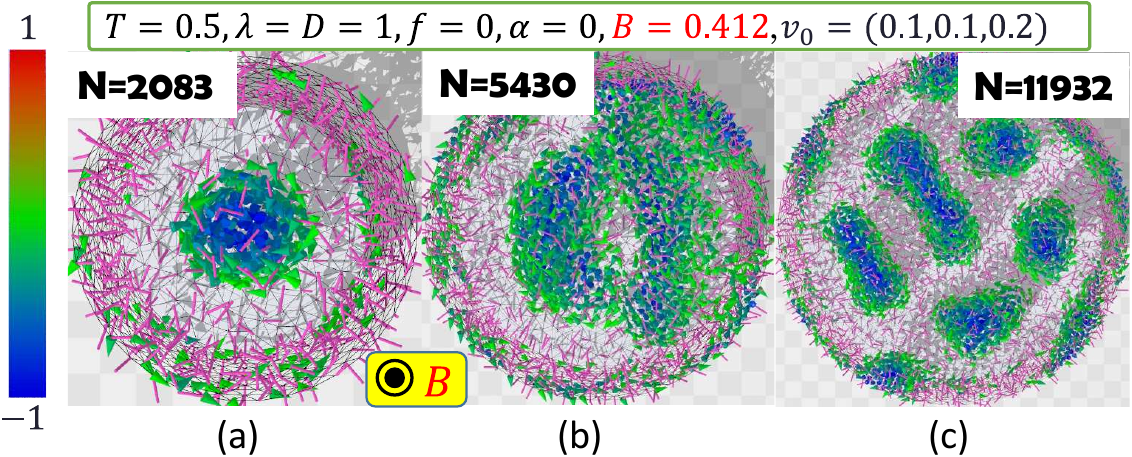}
	\caption{
		A confinement effect in which a stable sky on the lattice of (a) $N\!=\!2083$ becomes unstable on larger lattices (b) $N\!=\!5430$ and (c) $N\!=\!11932$. The spins of $\sigma^z\!<\!0$ are plotted. Small thin cylinders denote the $\vec{\tau}$ direction, which is isotropic because $f\!=\!0$. $T$ is the temperature ($k_B\!=\!1$ in the simulation unit).	\label{fig-2} }
\end{figure}
First, we show the effect of GC on sky stabilization observed experimentally in Ref. \cite{YWang-etal-NatCom2020} that sky confined in small nanodot is stable under a small magnetic field. To observe this GC effect, we numerically determine a set of parameters for the sky to be stable on the lattice of $N\!=\!2083$ and use the same parameters on the larger lattices of $N\!=\!5430$ and $N\!=\!11932$.  The $z$ component of spins is plotted in Figs. \ref{fig-2}(a)--(c), and we find that a stable sky on the $N\!=\!2083$ lattice becomes unstable as the lattice size increases. The spins of $\sigma^z\!<\!0$ are plotted.

\subsubsection{Strain Effect}
\begin{figure}[h]
	\centering{}\includegraphics[width=8.0cm,keepaspectratio,clip]{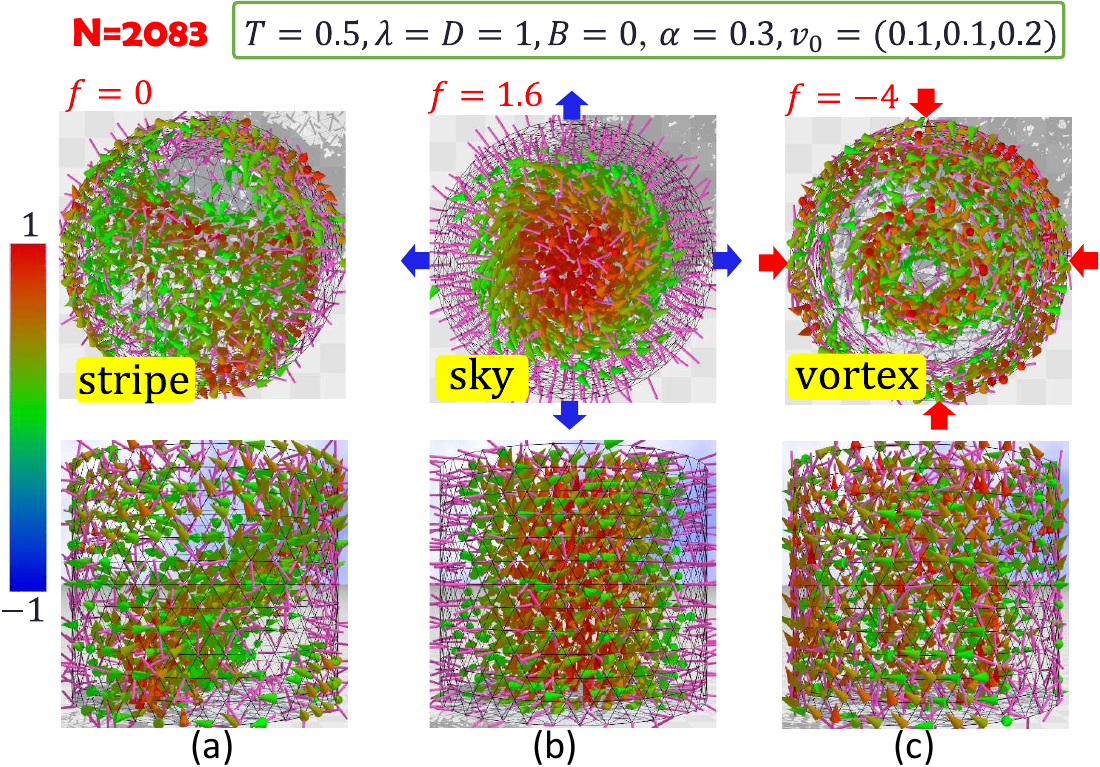}
	\caption{
		Top and side views of spin configurations obtained at (a) $f\!=\!0$,  (b) $f\!=\!1.6$, and (c) $f\!=\!-4$. The spins of $\sigma^z\!>\!0$ are plotted.		\label{fig-3} }
\end{figure}
We show the morphological changes in the spin configurations, including the sky, with the corresponding strain $\vec{\tau}$ configurations, under radial tension ($f\!>\!0$) and compression ($f\!<\!0$) with zero external magnetic field  ($B\!=\!0$).  In the case of $f\!=\!0$, a stripe phase appears (Fig. \ref{fig-3}(a)), and this stripe changes to sky when a tensile stress  $f\!=\!1.6$ is applied under $\alpha\!=\!0.3$ (Fig. \ref{fig-3}(b)), where the $\vec{\tau}$ is parallel to the radial direction as expected.  When a compression $f\!=\!-4$ applied under $\alpha\!=\!0.3$, a vortex configuration emerges  with a spiral $\vec{\tau}$ (Fig. \ref{fig-3}(c)). The numerically obtained morphological change in the spin configurations under a variation in $f$ is consistent with the experimentally reported results in Ref. \cite{YWang-etal-NatCom2020}. It is interesting to note that the spiral of $\vec{\tau}$ is accompanied by a vortex configuration of $\vec{\sigma}$ under a radial compression. The configurations including sky are stable though sky and stripe phases are not clearly separated. The spin direction at the center of sky is spontaneously determined in contrast to the case of skys with $B\!\not=\!0$ under $f\!=\!0$ shown in Fig. \ref{fig-2}.

\begin{figure}[h]
	\centering{}\includegraphics[width=8.0cm,keepaspectratio,clip]{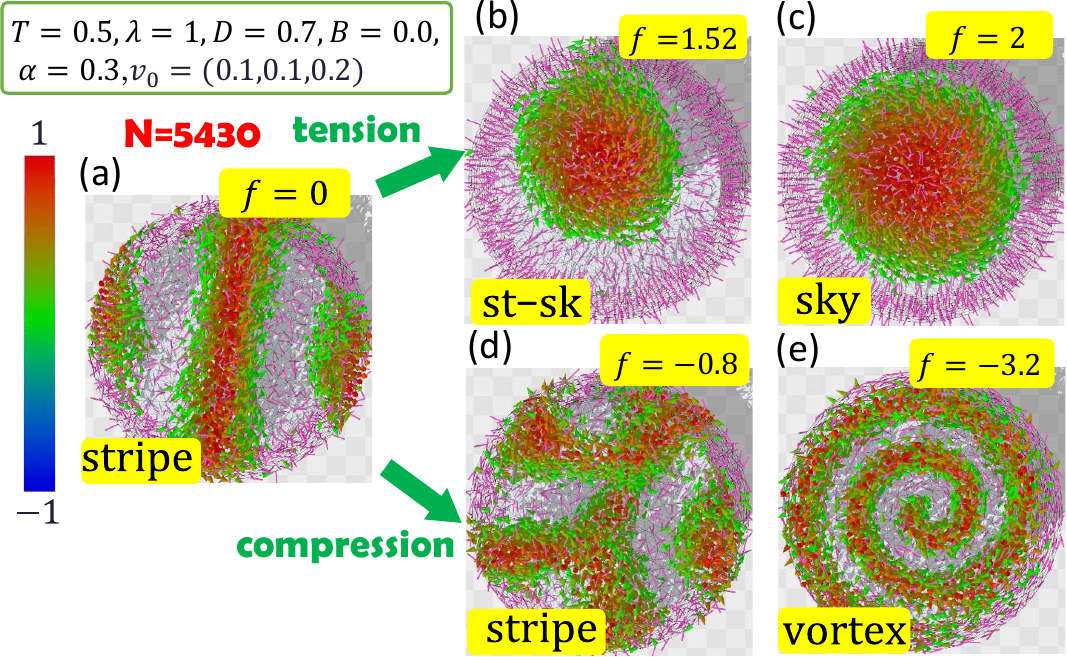}
	\caption{
		Morphological changes on the $N\!=\!5430$ lattice at $(T,\lambda,D)\!=\!(0.5,1,0.7)$ with  $B\!=\!0$. Snapshots of (a) $f\!=\!0$ (stripe), (b) $f\!=\!1.52$ (st-sk), (c) $f\!=\!2$ (skyrmion), (d) $f\!=\!-0.8$ (stripe) and (e) $f\!=\!-3.2$ (vortex). St-sk in (b) denotes an intermediate phase between stripe and sky. The spins of $\sigma^z\!>\!0$ are plotted. \label{fig-4} }
\end{figure}
To observe the strain effect in detail, we use a lattice of size $N\!=\!5430$, and plot the results in Figs. \ref{fig-4}(a)--(e) with the same parameters assumed on the $N\!=\!2083$ lattice except for $D\!=\!0.7$ (for sky size suitable to the lattice diameter). We find that the stripe phase at $f\!=\!\alpha\!=\!0$ changes to the st-sk, which is an intermediate phase between stripe and sky, and to the sky phase when the tensile stress $f(>0)$ increases to $f\!=\!1.52$, whereas it changes to the vortex phase when $f(<0)$ decreases. The stripe and vortex phases are now clear compared to the case of $N\!=\!2083$. 

\subsubsection{Direction-dependent Interaction Coefficients}
\begin{figure}[h]
	\centering{}\includegraphics[width=8.0cm,keepaspectratio,clip]{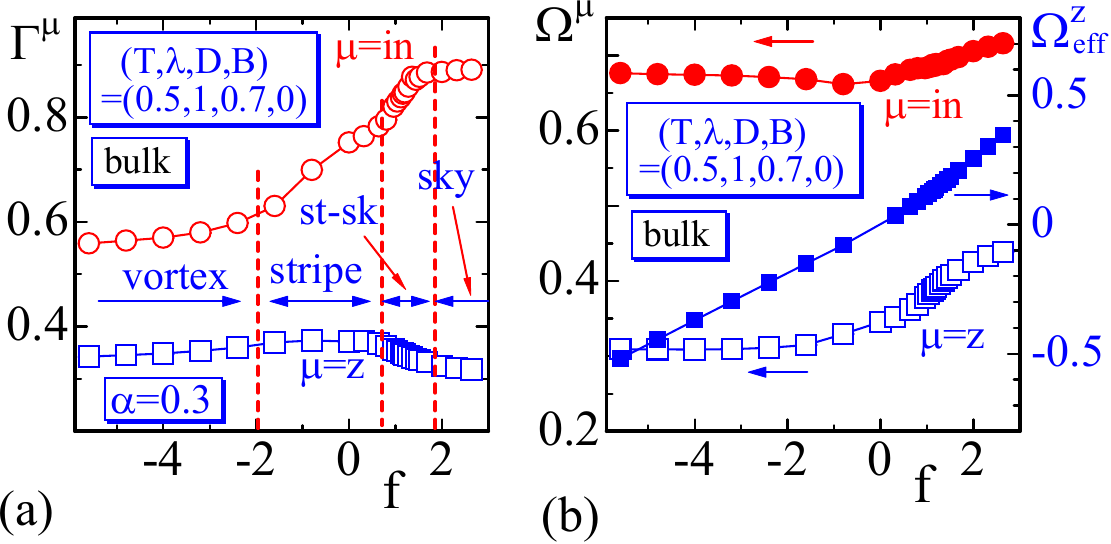}
	\caption{
		In-plane and $z$ direction components of (a) DMI and (b) ME coupling constants, where the in-plane components are defined by $\Gamma^{\rm in}\!=\!\Gamma^r\!+\!\Gamma^\theta$, $\Omega^{\rm in}\!=\!\Omega^r\!+\!\Omega^\theta$, and  $\Omega^z_{\rm eff}\!=\!\alpha f\Omega^z, (f\!\not=\!0)$ corresponding to  $K_{\rm av}$ in Ref.\cite{YWang-etal-NatCom2020}.   		\label{fig-5} }
\end{figure}
The effective coupling constants $\Gamma^{\rm in, z}$, $\Omega^{\rm in, z}$, and $\Omega^z_{\rm eff}\!=\!\alpha f\Omega^z, (\alpha\!=\!0.3, f\!\not=\!0)$,  are plotted in Figs. \ref{fig-5}(a),(b), where  $\Omega^z_{\rm eff}$ corresponds to the magnetic anisotropy $K_{\rm av}$ in Ref.\cite{YWang-etal-NatCom2020}. $\Gamma^{\rm in}$ increases when $f$  varies from negative (compression) to positive (tension) consistently with $D_{\rm av}$ reported in Ref.\cite{YWang-etal-NatCom2020}, where $D_{\rm av}$ decreases because the sign of the DMI energy in this paper is opposite to that in Ref.\cite{YWang-etal-NatCom2020}. The behavior that $\Omega^z_{\rm eff}$ increases with increasing $f$ from $f<0$ to $f>0$ is consistent with that of $K_{\rm av}$ in Ref.\cite{YWang-etal-NatCom2020}.

\begin{figure}[h]
	\centering{}
	\includegraphics[width=8.0cm,keepaspectratio,clip]{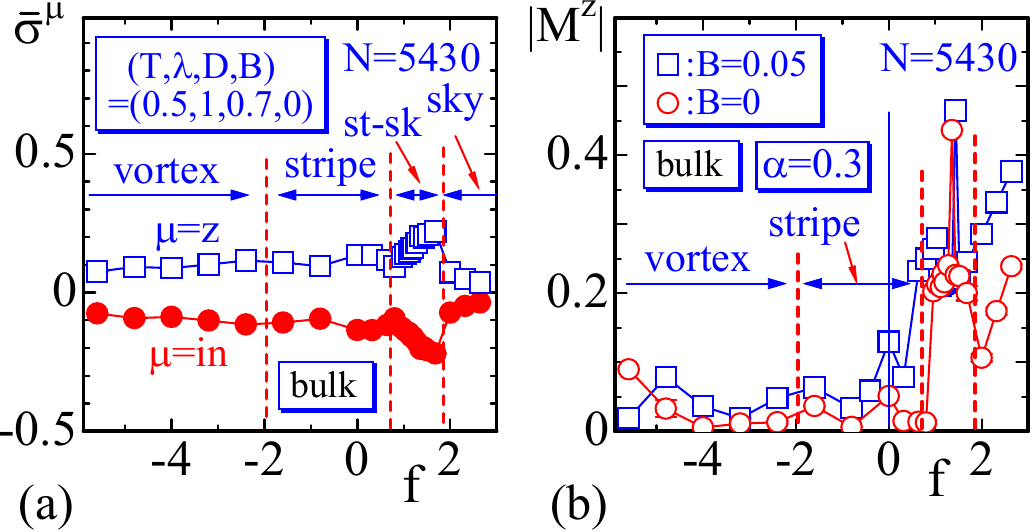}
	\caption{
		 (a) Non-polar order parameter $\bar\sigma^{z, \rm{in}}$ ($1\!\geq\!\bar\sigma^{z}\!\geq\!-0.5$)  and (b)  polar order parameter  $|M^z|$ ($1\!\geq\!|M^z|\!\geq\!0$).		\label{fig-6} }
\end{figure}

The nonpolar and polar order parameters $\bar{\sigma}^\mu$ and $|M^z|$ are  plotted (Figs. \ref{fig-6}(a),(b)), where $\bar\sigma^z\!=\!(3/2)(\langle(\sigma^z)^2\rangle\!-\!1/3)$, $\bar\sigma^{\rm in}\!=\!\bar\sigma^{x}\!+\!\bar\sigma^{y}\!=\!\bar\sigma^{r}\!+\!\bar\sigma^{\theta}$, and $|M^z| \!=\!|\sum_i\sigma_i^z|/\sum_i 1$. These $\bar{\sigma}^z$ and $\bar\sigma^{\rm in}$  in Fig. \ref{fig-6}(a) clarify  phase boundaries between the stripe, st-sk, and sky phases.  
We find from Fig. \ref{fig-6}(b) that $|M^z|\nearrow \left[f(>0)\nearrow\right]$, where $|M^z|$  for $B\!=\!0.05$ is included, and this behavior of $|M^z|$ is consistent with experimentally observed result that $|M^z|$ increases with increasing $f(>\!0)$ under external $B$ including $B\!=\!0$ at least in sky phase \cite{YWang-etal-NatCom2020}.  The sky range slightly increases in the $f$ axis when a small non-zero $B$ such as $B\!=\!0.05$ is applied.  


\section{Concluding Remarks}
We present tentative numerical results for geometrically confined (GC) skyrmions (skys) in nanodots simulated using a Finsler geometry (FG) model, in which anisotropies of interactions, including magnetic anisotropy, are dynamically generated by strains without spontaneous anisotropy. The results show that (i) the GC effect confines the sky in nanodots and stabilizes the sky in smaller nanodots with a small external magnetic field $(0,0,B)$. This GC effect originates from the surface effect, which makes the surface DMI smaller than the bulk DMI \cite{Diguet-etal-JMMM2023}. We also have (ii) radial strain effects that cause the sky to emerge in a steady state for tensile strain without external $B$. In addition to the stable sky states, an intermediate state between sky and stripe appears, and hence sky is not always clearly separated from the stripe phase. Further numerical studies are necessary. Detailed information on the models and numerical results will be reported elsewhere.

\noindent
\section*{Acknowledgments}
This work is supported in part by Collaborative Research Project J23Ly07 of the Institute of Fluid Science (IFS), Tohoku University. The numerical simulations were performed in part on the supercomputer system AFI-NITY at the Advanced Fluid Information Research Center, Institute of Fluid Science, Tohoku University.

\appendix
\noindent
{\bf Appendices}
\vspace{-0.2cm}

\section{Lattice Construction \label{App-A}}
\begin{figure}[h!]
	\centering{}\includegraphics[width=8.0cm,keepaspectratio,clip]{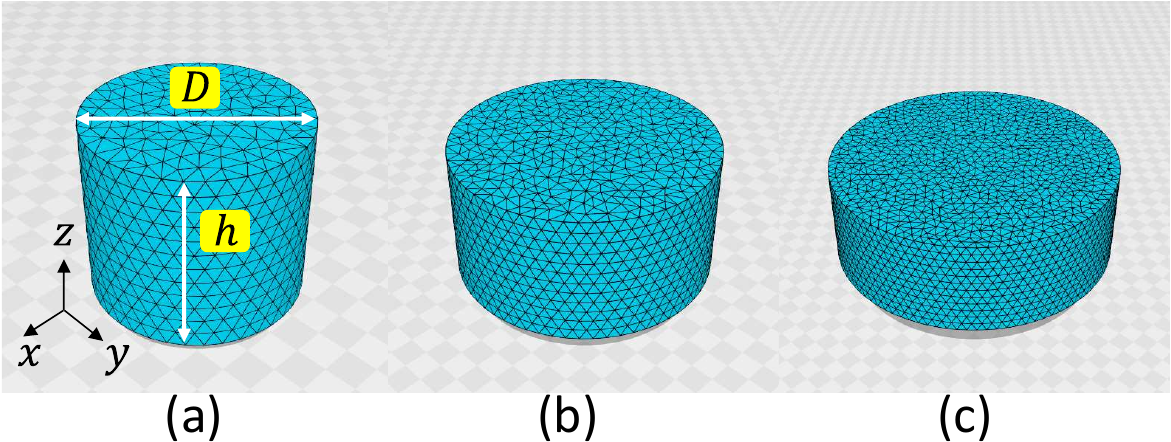}
	\caption{
		Cylindrical lattices of size (a) $N\!=\!2803$, (b) $N\!=\!5430$ and  (c) $N\!=\!11932$ composed of tetrahedrons.  The ratios $R\!=\!D/h$ of diameter $D$ and height $h$ are  $R\!=\!1.2,2,3$ in (a), (b) and (c), respectively, and proportional to $D$; 
		$h\!=\!12$ in the unit of the regular triangle height is fixed.		\label{fig-A-1} }
\end{figure}

Figure \ref{fig-A-1} shows the lattices for the simulations.

\section{Discretization of Direction-dependent Interaction Coefficient \label{App-B}}
\begin{figure}[b]
	\centering{}\includegraphics[width=8.0cm,keepaspectratio,clip]{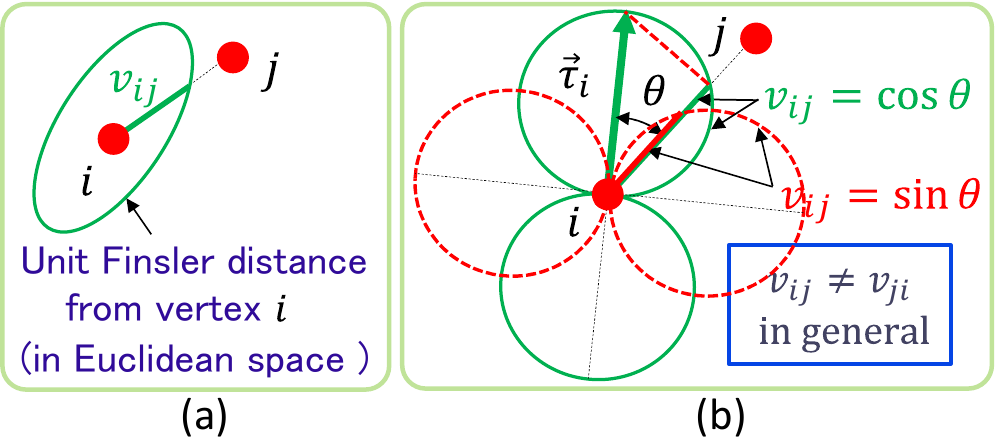}
	\caption{
		(a) An illustration of direction-dependent unit Finsler length $v_{ij}$ for interaction between particles $i$ and $j$, and (b) a graphical representation of the assumed $v_{ij}$ in the models (for $v_0\!=\!0$).		\label{fig-B-1} }
\end{figure}
The interaction coefficients in $H_{\rm FM, DM, ME}$ are given by
\begin{eqnarray}
	\label{Gamma-Omega-X}
	\begin{split}
		&\Gamma_{ij}=\frac{1}{\overline{n}}\bar{\Gamma}^{-1}\gamma_{ij}, \quad   \bar{\Gamma}=\frac{\sum_{\Delta}\sum_{ij(\Delta)}\langle \gamma_{ij}^{\rm{iso}}\rangle  }{\sum_{\Delta}\sum_{ij(\Delta)}1},\\
		&\Omega_{ij}=\frac{1}{\overline{n}}\bar{\Omega}^{-1}\omega_{ij}, \quad   \bar{\Omega}=\frac{\sum_{\Delta}\sum_{ij(\Delta)}\langle \omega_{ij}^{\rm{iso}}\rangle  }{\sum_{\Delta}\sum_{\Delta(ij))}1},
	\end{split}
\end{eqnarray}
where the symbol $\overline{n}\!=\!\sum_{ij}\sum_{\Delta (ij)}1/\sum_{ij}1$ denotes the mean value of $n_{ij}\!=\!\sum_{\Delta (ij)}1$, which is the total number of tetrahedrons $\Delta$ sharing bond $ij$. Here, we show only $\gamma_{12}, \omega_{12}$ and $\gamma_{13}, \omega_{13}$ on bonds 12 and 13 of the tetrahedron in Fig. \ref{fig-1}(a)
\begin{eqnarray}
	\label{Gammas}
	\begin{split}
		& \gamma_{12}=\frac{v_{12}}{v_{13}v_{14}}+\frac{v_{21}}{v_{23}v_{24}},\quad     \gamma_{13}=\frac{v_{13}}{v_{12}v_{14}}+\frac{v_{31}}{v_{32}v_{34}}, \\
		& \omega_{12}=\frac{v_{12}^3}{v_{13}v_{14}}+\frac{v_{21}^3}{v_{23}v_{24}},\quad     \omega_{13}=\frac{v_{13}^3}{v_{12}v_{14}}+\frac{v_{31}^3}{v_{32}v_{34}}, 
	\end{split}
\end{eqnarray}
where $\nu_{ij}$ is (Figs. \ref{fig-B-1}(a),(b))
\begin{eqnarray}
	\label{Finsler-unit-length}
		v_{ij}=\left\{ \begin{array}{@{\,}ll}
			|\vec{\tau}_i\cdot\vec{e}_{ij}| + v_0, \quad (v_0=0.1), \;(H_{\rm FM}, H_{\rm DM}) \\
			\sqrt{1-|\vec{\tau}_i\cdot\vec{e}_{ij}|^2} + v_0,\quad (v_0=0.2), \;(H_{\rm ME})
		\end{array} 
		\right.. 
\end{eqnarray}
$\langle \gamma_{ij}^{\rm{iso}}\rangle$ in $\Gamma_{ij}$ denotes the mean value of $\gamma_{ij}$ calculated from 1000 isotropic configurations of $\vec{\tau}$.  We should note that the mean value of $\Gamma_{ij}$  for isotropic configurations of  $\vec{\tau}$ satisfy $\langle \Gamma_{ij}\rangle\!\simeq\!1$. 
Due to this definition, $H_{FM}\!=\!\sum_{\Delta}\sum_{ij(\Delta)}\Gamma_{ij}\left(1-\vec{\sigma}_i\cdot\vec{\sigma}_j\right)$ for instance, returns to the standard one $\sum_{ij}\left(1-\vec{\sigma}_i\cdot\vec{\sigma}_j\right)$ for isotropic configurations of $\tau$.

\section{Decomposition of Interaction Coefficient and Corresponding Energies \label{App-C}}
\begin{figure}[h]
	\centering{}\includegraphics[width=8.0cm,keepaspectratio,clip]{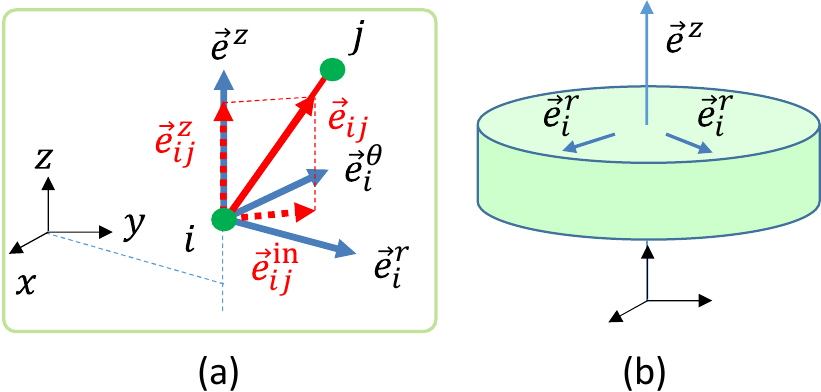}
	\caption{
		(a) Unit vector $\vec{e}_{ij}$ from vertices $i$ to $j$ can be decomposed into the in-plane $\vec{e}_{ij}^{\,\rm in}$  and $\vec{e}_{ij}^{\,z}$ components, (b) illustrations of the radial component $\vec{e}_i^{\,r}$ of $\vec{e}_{ij}$ at $i$. 	\label{fig-C-1} }
\end{figure}
Let $\vec{e}_i^{\,r}$, $\vec{e}_i^{\,\theta}$ and $\vec{e}^{\,z}$ be the unit vectors along the $r$, $\theta$ and $z$ directions at vertex $i$ (Figs. \ref{fig-C-1}(a),(b)). Then, we have a decomposition of $\vec{e}_{ij}$ such that $\vec{e}_{ij}\!=\!(\vec{e}_{ij}\cdot\vec{e}_i^{\,r})\vec{e}_i^{\,r}\!+\!(\vec{e}_{ij}\cdot\vec{e}_i^{\,\theta})\vec{e}_i^{\,\theta}\!+\!(\vec{e}_{ij}\cdot\vec{e}^{\,z})\vec{e}^{\,z}$. Using the expressions $1=\vec{e}_{ij}\cdot\vec{e}_{ij}\!=\!(\vec{e}_{ij}\cdot\vec{e}_i^{\,r})^2\!+\!(\vec{e}_{ij}\cdot\vec{e}_i^{\,\theta})^2\!+\!(\vec{e}_{ij}\cdot\vec{e}^{\,z})^2$ and $\Gamma_{ij}\!=\!\Gamma_{ij}\left((\vec{e}_{ij}\cdot\vec{e}_i^{\,r})^2\!+\!(\vec{e}_{ij}\cdot\vec{e}_i^{\,\theta})^2\!+\!(\vec{e}_{ij}\cdot\vec{e}^{\,z})^2\right)$, we have a decomposition of  $\Gamma_{ij}(\vec{\tau})$ such that
\begin{eqnarray}
	\label{Gamma-decomposition}
	\begin{split}
		&\Gamma_{ij}^{\rm in}(\vec{\tau})=\Gamma_{ij}(\vec{\tau})\left[(\vec{e}_{ij}\cdot\vec{e}_i^{\,r})^2+(\vec{e}_{ij}\cdot\vec{e}_i^{\,\theta})^2\right],\\
		&\Gamma_{ij}^{\,z}(\vec{\tau})=\Gamma_{ij}(\vec{\tau})(\vec{e}_{ij}\cdot\vec{e}^{\,z})^2.
	\end{split}
\end{eqnarray}
Using these direction dependent coefficients, the corresponding energies $H_{FM}$, $H_{DM}$, $H_{F2}$ can also be decomposed into direction dependent energies. Here, we show a decomposition of $H_{DM}$:
\begin{eqnarray}
	\label{HDMI-decomposition}
	\begin{split}
		&H_{DM}^{\rm in}=\frac{\sum_{\Delta}\sum_{ij(\Delta)}\Gamma_{ij}^{\rm in}(\vec{\tau})\vec{e}_{ij}\cdot \vec{\sigma}_i\times\vec{\sigma}_j}{\frac{1}{N_B}\sum_{\Delta}\sum_{ij(\Delta)}\Gamma_{ij}^{\rm in}(\vec{\tau})},\\
		&H_{DM}^{z}=\frac{\sum_{\Delta}\sum_{ij(\Delta)}\Gamma_{ij}^{z}(\vec{\tau})\vec{e}_{ij}\cdot \vec{\sigma}_i\times\vec{\sigma}_j}{\frac{1}{N_B}\sum_{\Delta}\sum_{ij(\Delta)}\Gamma_{ij}^{z}(\vec{\tau})},
	\end{split}
\end{eqnarray}
where $N_B\!=\!\sum_{ij}1$ is the total number of bonds. The denominators on the right hand side are the lattice averages of the components in Eq. (\ref{Gamma-decomposition}). Using these expressions, we have 
\begin{eqnarray}
	\begin{split}
		&H_{DM}=\sum_{\Delta}\sum_{ij(\Delta)}\Gamma_{ij}(\vec{\tau})\vec{e}_{ij}\cdot \vec{\sigma}_i\times\vec{\sigma}_j\\
		=&\sum_{\Delta}\sum_{ij(\Delta)}\left(\Gamma_{ij}^{\rm in}+\Gamma_{ij}^{\,z}\right)\vec{e}_{ij}\cdot \vec{\sigma}_i\times\vec{\sigma}_j \\
		=&\left(\frac{1}{N_B}\sum_{\Delta}\sum_{ij(\Delta)}\Gamma_{ij}^{\rm in}\right) H_{DM}^{\rm in}+
		\left(\frac{1}{N_B}\sum_{\Delta}\sum_{ij(\Delta)}\Gamma_{ij}^{\,z}\right) H_{DM}^{z},
	\end{split}
\end{eqnarray}
and therefore, the mean value is given by
\begin{eqnarray}
	\label{effective-HDMI}
	\begin{split}
		\langle H_{DM}\rangle=\langle\Gamma_{ij}^{\rm in}\rangle \langle H_{DM}^{\rm in}\rangle+\langle\Gamma_{ij}^{\,z}\rangle \langle H_{DM}^{z}\rangle,
	\end{split}
\end{eqnarray}
where the mean value is the sample or ensemble average calculated via the MC simulations. The expression of $H_{DM}$ in Eq. (\ref{effective-HDMI}) indicates that the direction dependent DMI coefficients in Eq.(\ref{Gamma-decomposition}) and the corresponding DMI energies in Eq.(\ref{HDMI-decomposition}) are reasonable.

\end{document}